\documentclass[aps,pra,reprint,superscriptaddress,nofootinbib]{revtex4-1}
\usepackage{amsmath}
\usepackage{amssymb}
\usepackage{mathrsfs}
\usepackage{graphicx}
\usepackage[caption=false]{subfig}
\usepackage{enumitem}
\usepackage[normalem]{ulem}
\usepackage[percent]{overpic}
\usepackage{bm}
\usepackage{rotating}
\usepackage[usenames, dvipsnames]{color}
\usepackage{hyperref}
\usepackage{cancel}
\usepackage{verbatim}
\usepackage{mathtools}
\usepackage{graphicx}
\usepackage{tensor}
\usepackage{verbatim}
\usepackage{MnSymbol}
\usepackage{booktabs}
\graphicspath{ {images/} }

\usepackage{braket}

\newcommand{\ii}{\mathrm{i}}

\renewcommand{\d}{\mathrm{d}}
\renewcommand{\vec}{\text{vec}}

\newcommand{\be}{\begin{equation}}
\newcommand{\bel}[1]{\begin{equation}\label{#1}}
\newcommand{\ee}{\end{equation}}

\begin{document}
\title{Gaussian ancillary bombardment}
\author{Daniel Grimmer}
\email{dgrimmer@uwaterloo.ca}
\affiliation{Institute for Quantum Computing, University of Waterloo, Waterloo, ON, N2L 3G1, Canada}
\affiliation{Dept. Physics and Astronomy, University of Waterloo, Waterloo, ON, N2L 3G1, Canada}

\author{Eric Brown}
\email{ericgb86@gmail.com}
\affiliation{ICFO-Institut de Ciencies Fotoniques, The Barcelona Institute of Science and Technology, 08860 Castelldefels (Barcelona), Spain}

\author{Achim Kempf}
\affiliation{Dept. Applied Math., University of Waterloo, Waterloo, ON, N2L 3G1, Canada}
\affiliation{Dept. Physics and Astronomy, University of Waterloo, Waterloo, ON, N2L 3G1, Canada}
\affiliation{Institute for Quantum Computing, University of Waterloo, Waterloo, ON, N2L 3G1, Canada}
\affiliation{Perimeter Institute for Theoretical Physics, Waterloo, ON, N2L 2Y5, Canada}

\author{Robert B. Mann}
\email{rbmann@uwaterloo.ca}
\affiliation{Dept. Physics and Astronomy, University of Waterloo, Waterloo, ON, N2L 3G1, Canada}
\affiliation{Institute for Quantum Computing, University of Waterloo, Waterloo, ON, N2L 3G1, Canada}
\affiliation{Perimeter Institute for Theoretical Physics, Waterloo, ON, N2L 2Y5, Canada}

\author{Eduardo Mart\'{i}n-Mart\'{i}nez}
\email{emartinmartinez@uwaterloo.ca}
\affiliation{Institute for Quantum Computing, University of Waterloo, Waterloo, ON, N2L 3G1, Canada}
\affiliation{Dept. Applied Math., University of Waterloo, Waterloo, ON, N2L 3G1, Canada}
\affiliation{Perimeter Institute for Theoretical Physics, Waterloo, ON, N2L 2Y5, Canada}

\begin{abstract}
We analyze in full detail the time evolution of an open Gaussian quantum system rapidly bombarded by Gaussian ancillae. As a particular case this analysis covers the thermalization (or not) of a harmonic oscillator coupled to a thermal reservoir made of harmonic oscillators. We derive general results for this scenario and apply them to the problem of thermalization.  We show that only a particular family of system-environment couplings will cause the system to thermalize to the temperature of its environment. We discuss that if we want to understand thermalization as ensuing from the Markovian interaction of a system with the individual microconstituents of its (thermal) environment then the process of thermalization is not as robust as one might expect.
\end{abstract}

\maketitle
\section{Introduction}
Open quantum dynamics---the study of the evolution of quantum systems interacting with an environment---has wide sweeping theoretical and experimental importance. It is fundamental in the study of quantum thermodynamics. Since thermalization is a non-unitary process, it requires an environment. Open dynamics is also critical in understanding the noise and decoherence modes ubiquitously present in experimental settings \cite{Nielsen:2000}. 

The formalism of Gaussian Quantum Mechanics (GQM), (see, e.g., \cite{GQMRev}) simplifies the treatment of many quantum mechanical problems by making use of the phase space representation of quantum mechanics, focusing on states that can be fully characterized with a Gaussian Wigner function. Such states are theoretically and experimentally relevant, including coherent states, thermal states, and squeezed states. As long as all the relevant transformations preserve this Gaussianity (i.e. take Gaussian states to Gaussian states), GQM provides a significant decrease in the overhead of describing quantum states and transformations. One needs only track the system's first and second statistical moments instead of a vector in an infinite dimensional Hilbert space. The literature abounds with reviews on Gaussian quantum mechanics, in particular in its applications to quantum information; the reader is referred to \cite{weedbrook, adesso1, lami}.

In this paper, we consider the dynamics induced in a generic Gaussian system when rapidly bombarded by a series of Gaussian ancillae, a scenario we call \textit{Gaussian ancillary bombardment}. An intuitive example of such a scenario is a harmonic oscillator in a thermal bath of harmonic oscillators. 

To study the general scenario, in Sec. \ref{InterpolateGQM} we adapt the rapid repeated interaction formalism developed in \cite{Grimmer2016a,Grimmer2017a} to the Gaussian setting. Specifically, we construct an interpolating master equation for the discrete time dynamics induced by the rapid interactions. In Sec. \ref{AncillaryBombardmentGQM} we apply this adapted formalism to the a generic Gaussian ancillary bombardment scenario and analyze the resulting master equation. In this analysis, we make use of the partition of open Gaussian dynamics developed in \cite{ArXivGrimmer2017b} to characterize the dynamics in terms of unitarity, ability to cause energy flow, state-dependence and mode mixing. 

Finally, in Sec. \ref{Example}, we apply the tools built in this paper to the problem of understanding thermalization as resulting from the Markovian bombardment of a small system by the microconstituents of a thermal reservoir. We show that if we are to model equilibration and thermalization as resulting from this kind of dynamics then these processes critically depend on the system-environment coupling.

The methods and results we present not only add to a growing understanding of Gaussian open dynamics \cite{koga, nicacio, nicacio2} but also provide tools for investigating 
the thermodynamics of systems that are repeatedly disturbed by an environment, particularly with regard to microscopic details connected with the flow of energy and information.

\section{Gaussian Quantum Mechanics}\label{ReviewGQM}

Consider a system composed of $N$ coupled modes (for example, harmonic oscillators) with the $n^{th}$ of these modes characterized by its quadrature operators, $\hat{q}_n$ and $\hat{p}_n$, which obey the canonical bosonic commutation relations,
\be
[\hat{q}_n,\hat{q}_m]
=[\hat{p}_n,\hat{p}_m]
=0
\quad\text{and}\quad
[\hat{q}_n,\hat{p}_m]=\ii \,  \delta_{nm} \, \hat{\openone}.
\ee

Such systems can be fully described in terms of a pseudo-probability distribution defined on the system's phase space \cite{Groenewold,Moyal}. In particular, a state with density matrix $\rho$ can be equivalently represented by its Wigner function,
\be
W(\bm{q},\bm{p})=\frac{1}{\pi^N}\!\int_{-\infty}^\infty \d^N \bm{s}
\bra{\bm{q}+\bm{s}}\rho\ket{\bm{q}-\bm{s}}\exp(-2\ii \, \bm{p}\cdot\bm{s}).
\ee

Gaussian Quantum Mechanics (GQM) is the restriction of quantum mechanics to the class of states whose Wigner functions are Gaussian and to the class of transformations which preserve this Gaussianity. The following summary of GQM significantly summarizes the in-depth summary given in \cite{ArXivGrimmer2017b} in which many of the following claims are spelled out and demonstrated. 

The main benefit of this restriction to Gaussian states and transformations is that it allows for a significantly simplified description of quantum states and transformations whilst still describing a wide variety of theoretically and experimentally relevant situations. In particular, a Gaussian distribution is  completely determined by its first and second statistical moments. Thus collecting the system's quadrature operators into the vector
\bel{XhatDef}
\hat{\bm{X}}
\coloneqq
(\hat{q}_1,\hat{p}_1,\hat{q}_2,\hat{p}_2,\dots,\hat{q}_N,\hat{p}_N)^\intercal,
\ee
a Gaussian state is fully described by (a) the mean of each of these operators, collected in the $2N$-dimensional mean vector
\bel{XDef}
\bm{X}
\coloneqq\langle\hat{\bm{X}}\rangle
=\big(\langle\hat{q}_1\rangle,\langle\hat{p}_1\rangle,\dots,\langle\hat{q}_N\rangle,\langle\hat{p}_N\rangle\big)^\intercal,
\ee
and (b) by the covariances between them, collected in the the $2N$ by $2N$ symmetric covariance matrix
\bel{Vdef}
\sigma_j{}^k
\coloneqq
\big\langle
\hat{X}_j \, \hat{X}^k
+ \hat{X}^k \, \hat{X}_j
\big\rangle
-2\big\langle\hat{X}_j\big\rangle
\big\langle\hat{X}^k\big\rangle.
\ee

Note that any two quadrature operators, say $\hat{X}_j$ and $\hat{X}^k$, will either commute to $\ii \,  \hat{\openone}$ or to $0$ such that all of the system's commutation relations are captured by the phase space matrix $\Omega$, defined as 
\begin{align}\label{OmegaDef}
[\hat{X}_j,\hat{X}^k]
&=\ii \ \Omega_j{}^k \, \hat{\openone}.
\end{align}
This matrix, called the symplectic form, is given explicitly as
\bel{OmegaExplicit}
\Omega
=\bigoplus_{n=1}^N \omega
=\openone_N\otimes\omega; \ \ \ \ \omega
=\begin{pmatrix}
0 & 1\\
-1 & 0
\end{pmatrix},
\ee
in the same representation as \eqref{XhatDef}. Note that $\Omega$ is real-valued, antisymmetric, and invertible with \mbox{$\Omega^{-1}=\Omega^T=-\Omega$}. 

As in standard quantum mechanics, in GQM the commutation relations underlie the uncertainty principle, which all valid states obey. For Gaussian states the uncertainty principle is \cite{Simon1994},
\bel{SigmaPosCond}
\sigma\geq\ii \, \Omega.
\ee
For a matrix $M$, the notation $M\geq 0$ indicates here that $M$ is positive semi-definite. Moreover $M_1\geq M_2$ here means $M_1-M_2\geq0$. The uncertainty bound \eqref{SigmaPosCond} implies that that $\sigma\geq0$ (see Sec. II in \cite{ArXivGrimmer2017b}).

Gaussian unitary transformations are unitary transformations in the system's Hilbert space that preserve the Gaussianity of the state. Differential Gaussian unitary transformations are generated by Hamiltonians that  are at most quadratic in the the operator vector \cite{Schumaker1986}. Such Hamiltonians can always be cast in the form,
\bel{QuadHamForm}
\hat{H}=\frac{1}{2}\hat{\bm{X}}^\intercal F \, \hat{\bm{X}}
+\bm{\alpha}^\intercal\hat{\bm{X}}.
\ee
where $F$ is a $2N$ by $2N$ real symmetric matrix and $\bm{\alpha}$ is a real-valued $2N$ dimensional vector. From \eqref{QuadHamForm}, one can calculate the evolution of the mean vector, $\bm{X}$, and of the covariance matrix, $\sigma$, as
\begin{align}
\label{SymplecticDiffXUpHam}
\frac{\d}{\d t}\bm{X}(t)
&=\Omega (F \bm{X}(t)+\bm{\alpha}),\\
\label{SymplecticDiffVUpHam}
\frac{\d}{\d t}\sigma(t)
&=(\Omega \, F) \, \sigma(t)
+\sigma(t) \, (\Omega \, F)^\intercal.
\end{align}
For a time-independent Hamiltonian, integrating these equations for a time interval $[0,t]$ gives
\begin{align}
\label{SymplecticXUp}
\bm{X}(0)&\longrightarrow \bm{X}(t)=S(t) \, \bm{X}(0)+\bm{d}(t),\\
\label{SymplecticVUp}
\sigma(0)&\longrightarrow \sigma(t)=S(t) \, \sigma(0) \, S^\intercal(t)
\end{align}
where
\begin{align}
\label{SHamDef}
S(t)&=\text{exp}(\Omega F \, t),\\
\label{dHamDef}
\bm{d}(t)&=\frac{\text{exp}(\Omega F \, t)-\openone_{2N}}{\Omega F} \, \Omega\bm{\alpha}.
\end{align}
Note that \eqref{dHamDef} does not require $\Omega F$ to be invertible. Instead the notation can be understood in terms of the following series expansion
\bel{(ExpX-1)byXDef}
\frac{\text{exp}(X \, t)-\openone}{X}
=\sum_{m=0}^\infty \frac{t^{m+1}}{(m+1)!}X^m.
\ee

More generally, any transformation of the form \eqref{SymplecticXUp} and \eqref{SymplecticVUp} (i.e., with generic $S$ and $\bm{d}$) can be implemented by evolving under a (potentially time dependent\footnote{Notice that in order to implement a general sympletic transformation a time dependent generator is generally needed. This follows from the exponential in the symplectic group not being surjective.}) quadratic Hamiltonian with the sole restriction that it preserves the symplectic form (i.e., the commutation relation) as 
\bel{SympTranDef}
S \, \Omega \, S^\intercal=\Omega.
\ee 
Such a matrix $S$ implements a symplectic transformation. Together with $\bm{d}$, the update \eqref{SymplecticXUp} and \eqref{SymplecticVUp} constitutes a symplectic-affine transformation. Gaussian unitary transformations on the system's Hilbert space correspond to symplectic-affine transformations on the system's phase space.

In addition to the Gaussian unitary transformations described above, one can implement non-unitary Gaussian transformations by allowing the system to interact with an environment. In direct analogy with the Stinespring dilation theorem, one can implement any completely positive trace preserving (CPTP) Gaussian transformation as a Gaussian unitary transformation in some larger Hilbert space (or equivalently as a symplectic-affine transformation in a larger phase space) \cite{GaussianDilation}. From this it follows that the most general form of Gaussian update on $\bm{X}$ and $\sigma$ is,
\begin{align}
\label{GeneralUpdateX}
\bm{X}(0)&\to \bm{X}(t)=T(t)\bm{X}(0)+\bm{d}(t),\\
\label{GeneralUpdateV}
\sigma(0)&\to \sigma(t)=T(t) \, \sigma(0) \, T^\intercal(t)+R(t).
\end{align}
where $\bm{d}(t)$ is a real $2N$-dimensional vector, $T(t)$ and $\bm{R}(t)$ are $2N$ by $2N$ real matrices, $R(t)$ is symmetric, and $T(t)$ (unlike $S$) is not necessarily symplectic.

A transformation (given by $T$, $\bm{d}$, $R$) is CPTP if and only if it obeys the complete positivity condition \cite{GQMRev}
\bel{FiniteCPCond}
R\geq\ii \, (T \, \Omega \,  T^\intercal-\Omega).
\ee
where a sketch of the proof appears in the appendix of \cite{ArXivGrimmer2017b}. Recall the notation $M\geq 0$ indicates that $M$ is a positive semi-definite matrix.

We can take the update given by \eqref{GeneralUpdateX} and \eqref{GeneralUpdateV} to be differential, as
\begin{align}
T(\d t)&=\openone_{2N}+\d t \ \Omega  \, A,\\
\bm{d}(\d t)&=\d t \ \Omega \, \bm{b},\\
R(\d t)&=\d t \ C,
\end{align}
where $\bm{b}$ is a real $2N$-dimensional vector, $A$ and $C$ are $2N$ by $2N$ real matrices, $C$ is symmetric. Since $\Omega$ is invertible, and since $A$ and $\bm{b}$ are arbitrary, assuming that a factor of $\Omega$ precedes $A$ and $\bm{b}$ is justified.

From this differential update one can find that the general form of the Gaussian master equations is
\begin{align}
\label{GeneralDiffXUp}
\frac{\d}{\d t}\bm{X}(t)
&=\Omega(A(t) \bm{X}(t)+\bm{b}(t)),\\
\label{GeneralDiffVUp}
\frac{\d}{\d t}\sigma(t)
&=(\Omega A(t)) \, \sigma(t)
+\sigma(t) \, (\Omega A(t))^\intercal
+C(t).
\end{align}
The differential version of the complete positivity condition \eqref{FiniteCPCond} is
\bel{DiffCPCond}
C\geq\ii \, \Omega (A-A^\intercal)\Omega
\ee
from which it follows that $C\geq0$.

In \cite{ArXivGrimmer2017b} the dynamical effect of the $A$, $\bm{b}$, and $C$ terms were explored in detail. To summarize, the effect of the $A$ term is to implement rotations, squeezings, and amplifications in phase space, whereas the $\bm{b}$ term implements displacement and the $C$ term implements state-independent noise.

For time-independent generators ($A$, $\bm{b}$, and $C$), integrating these equations for a time interval $[0,t]$ gives an update of the form \eqref{GeneralUpdateX} and \eqref{GeneralUpdateV}  with
\begin{align}
\label{TfromAbC}
T(t)&=\exp(\Omega A  \, t),\\
\label{dfromAbC}
\bm{d}(t)&=\frac{\exp(\Omega A \,  t)-\openone_{2N}}{\Omega A} \, \Omega \, \bm{b},\\
\label{RfromAbC}
R(t)&=\text{vec}^{-1}\Big(\frac{\exp((\Omega A\otimes\Omega A) \,  t)-\openone_{4N^2}}{\Omega A\otimes\Omega A} \ \text{vec}(C)\Big).
\end{align}
where the $\vec$ operation is defined \cite{ArXivGrimmer2017b} to map outer products to tensor products as
\bel{OuterToTensor}
\vec(\lambda \ \bm{u}\bm{v}^\intercal)
\coloneqq\lambda \ \bm{u}\otimes\bm{v}
\ee
for some scalar $\lambda$ and vectors $\bm{u}$ and $\bm{v}$. By linearity this defines its action on any matrix. 
%\gr{[A minor point, but I feel like this is a messy way to introduce the vec operation. It may be better to start with definining the operation and then discussing its properties. Whichever way you guys feel is better.]} \tcr{\bf [Dan: I like it this way, it explains quickly where the tensor products in \eqref{RfromAbC} come from]} {\color{BurntOrange} \bf [Edu: I also like it as is, the definition as an abstract map and a representaiton later I believe is the best way to present it.]} 
One quickly finds that for any matrices $X$, $Y$ and $Z$
\bel{VecIdentity}
\vec(X \, Y \, Z^\intercal)=(X\otimes Z)\vec(Y).
\ee
This operation can be represented by the vector formed by taking the entries of a matrix in order as follows,
\be
\vec\begin{pmatrix}
a & b \\
c & d
\end{pmatrix}
=(a,b,c,d)^\intercal.
\ee
Note that $\text{vec}^{-1}$ is trivially defined by ``restacking'' the matrices entries.

Also, as before, note that it is not necessary that $\Omega A$ and $\Omega A\otimes\Omega A$ are invertible for us to evaluate \eqref{dfromAbC} and \eqref{RfromAbC} as we can make use of the series \eqref{(ExpX-1)byXDef}.

\section{Rapid Repeated Gaussian Interaction}\label{InterpolateGQM}
In this section we build a Gaussian master equation of the general form \eqref{GeneralDiffXUp} and \eqref{GeneralDiffVUp} from rapid repeated application of a Gaussian channel of the general form \eqref{GeneralUpdateX} and \eqref{GeneralUpdateV}.

Specifically, we take a Gaussian system (characterized by its mean vector, $\bm{X}$, and its covariance matrix, $\sigma$) to be updated in discrete time steps of duration $\delta t$ via the Gaussian channel given by some $T(\delta t)$, $\bm{d}(\delta t)$, and $R(\delta t)$ as
\begin{align}
\label{UpdateSchemeXX}
\bm{X}((n+1)\delta t)
&=T(\delta t) \, \bm{X}(n \, \delta t)
+\bm{d}(\delta t),\\
\label{UpdateSchemeVV}
\sigma((n+1)\delta t)
&=T(\delta t) \, \sigma(n \, \delta t) \, T^\intercal(\delta t)
+R(\delta t).
\end{align}
Given the initial system state, $\bm{X}(0)$ and $\sigma(0)$, the above update scheme defines the system state at the discrete time points $t=n\,\delta t$. Note this update is Markovian since it is time-local (it only depends on the current state of the system).

Further we make the natural assumptions that 
\bel{NothingNoTime}
T(0)=\openone_{2N}, \ \ \  \bm{d}(0)=0, \ \ \ \text{and} \ \ \ R(0)=0
\ee
(nothing happens in no time) and that 
\bel{FiniteRate}
T'(0), \ \ \  \bm{d}'(0), \ \ \ \text{and} \ \ \ R'(0) \ \ \ \text{exist}
\ee
(things happen at a finite rate). Finally we assume that the update scheme is invertible. Ultimately, this means that $T(\delta t)$ is non-singular. Note that we automatically have this for small enough $\delta t$.

From the above update we seek to construct a Gaussian master equation of the general form
\begin{align}
\label{GQMInterpMasterEqsX}
\bm{X}'(t)
&=\Omega(A_{\delta t} \, \bm{X}(t)
+\bm{b}_{\delta t}),\\
\label{GQMInterpMasterEqsV}
\sigma'(t)
&=(\Omega A_{\delta t}) \, \sigma(t)
+\sigma(t) \, (\Omega A_{\delta t})^\intercal
+C_{\delta t}
\end{align}
for some generators $A_{\delta t}$, $\bm{b}_{\delta t}$, and $C_{\delta t}$ such that the dynamics it describes exactly matches the dynamics given by the discrete updater at every time point, $t=n \, \delta t$. As the dynamics generated by \eqref{GQMInterpMasterEqsX} and \eqref{GQMInterpMasterEqsV} is defined for all $t\geq0$ (not just $t=n\,\delta t$) this master equation constitutes an interpolation scheme (see \cite{Grimmer2016a} for details). 

In general, such an interpolation scheme is not uniquely determined. However, as discussed in \cite{Grimmer2017a}, there is a unique interpolation scheme with time-independent generators which converge in the rapid interaction limit (as $\delta t\to0$).

This unique interpolation scheme is constructed in detail in Appendix \ref{AppGQMInterpolate}, yielding the interpolation generators
\begin{align}
\label{AdtDef}
\Omega A_{\delta t}
&=\frac{1}{\delta t}\text{Log}(T(\delta t)),\\
\label{bdtDef}
\Omega \, \bm{b}_{\delta t}
&=\frac{1}{\delta t}
\frac{\text{Log}(T(\delta t))}{T(\delta t)-\openone_{2N}}\bm{d}(\delta t),\\
\label{CdtDef}
C_{\delta t}
&=\vec^{-1}\Big(\frac{1}{\delta t}\frac{\text{Log}(T(\delta t) \otimes T(\delta t))}{T(\delta t) \otimes T(\delta t)-\openone_{4N^2}} \, \vec\big(R(\delta t)\big)\Big).
\end{align}
where we emphasize that
the expressions for $\bm{b}_{\delta t}$, and $C_{\delta t}$ are to be understood  via the series expansion
\bel{LogSeries2}
\frac{\text{Log}(X)}{X-\openone}
=\sum_{m=0}^\infty\frac{(-1)^m}{m+1}(X-\openone)^m
\ee
and so  \mbox{$T(\delta t)-\openone_{2N}$} and \mbox{$T(\delta t)\, \otimes \, T(\delta t)-\openone_{4N^2}$} need not be invertible.
Finally, we note that in the above equations we take the logarithm's principle branch cut, such that $\text{Log}(\openone)=0$. This assures that the interpolation generators converge as $\delta t\to0$.

If in addition to the minimal regularity assumed above --- \eqref{NothingNoTime} and \eqref{FiniteRate} --- we have that $T(\delta t)$, $\bm{d}(\delta t)$, and $R(\delta t)$ are analytic at $\delta t=0$, then we can then expand them as a series in $\delta t$ as
\begin{align}
\label{TSeries}
T(\delta t)
&=\openone_{2N}
+\delta t \, T_1
+\delta t^2 \, T_2
+\delta t^3 \, T_3
+\delta t^4 \, T_4
+\dots,\\
\label{dSeries}
\bm{d}(\delta t)
&=0
+\delta t \, \bm{d}_1
+\delta t^2 \, \bm{d}_2
+\delta t^3 \, \bm{d}_3
+\delta t^4 \, \bm{d}_4
+\dots,\\
\label{RSeries}
R(\delta t)
&=0
+\delta t \, R_1
+\delta t^2 \, R_2
+\delta t^3 \, R_3
+\delta t^4 \, R_4
+\dots \, .
\end{align}
Using these series expansions, through \eqref{AdtDef}, \eqref{bdtDef}, and \eqref{CdtDef}, we can expand each interpolation generator as a series in $\delta t$ as well,
\begin{align}
\label{ASeries}
A_{\delta t}
&=A_0
+\delta t \, A_1
+\delta t^2 \, A_2
+\delta t^3 \, A_3
+\dots,\\
\label{bSeries}
\bm{b}_{\delta t}
&=\bm{b}_0
+\delta t \, \bm{b}_1
+\delta t^2 \, \bm{b}_2
+\delta t^3 \, \bm{b}_3
+\dots,\\
\label{CSeries}
C_{\delta t}
&=C_0
+\delta t \, C_1
+\delta t^2 \, C_2
+\delta t^3 \, C_3
+\dots,
\end{align}
where the first few terms of the expansion of $A_{\delta t}$ are given by
\begin{align}
\label{A0def}
\Omega A_0=T_1,&\\
\label{A1def}
\Omega A_1=T_2
&-\frac{1}{2}T_1{}^2,\\
\label{A2def}
\Omega A_2=T_3
&-\frac{1}{2}(T_1 T_2+T_2 T_1)
+\frac{1}{3}T_1{}^3.
%\label{A3def}
%\Omega A_3=T_4
%&-\frac{1}{2}(T_1 T_3+T_2 T_2+T_3 T_1)\\
%\nonumber
%&+\frac{1}{3}(T_1{}^2 T_2+T_2 T_1{}^2+T_1 T_2 T_1)
%-\frac{1}{4}T_1{}^4.
\end{align}
The first few terms of the expansion of $\bm{b}_{\delta t}$ are given by
\begin{align}
\Omega \, \bm{b}_0
=\bm{d}_1&,\\
\Omega \, \bm{b}_1
=\bm{d}_2
&-\frac{1}{2}T_1\bm{d}_1,\\
\Omega \, \bm{b}_2
=\bm{d}_3
&-\frac{1}{2}(T_1\bm{d}_2+T_2\bm{d}_1)
+\frac{1}{3}T_1^2\bm{d}_1.
%\Omega \, \bm{b}_3
%=\bm{d}_4
%&-\frac{1}{2}(T_1\bm{d}_3+T_2\bm{d}_2+T_3\bm{d}_1)\\
%\nonumber
%&+\frac{1}{3}(T_1^2\bm{d}_2+T_2 T_1\bm{d}_1+T_1 T_2\bm{d}_1)
%-\frac{1}{4}T_1^3\bm{d}_1.
\end{align}
Finally, the first few terms of the expansion of $C_{\delta t}$ are given by
\begin{align}
C_0&=R_1,\\
C_1&=R_2 
-\frac{1}{2}(T_1 R_1+R_1 T_1^\intercal),\\
C_2&=R_3 
-\frac{1}{2}(T_2 R_1+R_1 T_2^\intercal+T_1 R_2+R_2 T_1^\intercal)\\
\nonumber
&+\frac{1}{3}(T_1{}^2 R_1+R_1 T_1^\intercal{}^2)
+\frac{1}{6} T_1 R_1 T_1^\intercal.
%C_3&=R_4\\
%\nonumber
%&-\frac{1}{2}
%(T_3 R_1+R_1 T_3^\intercal
%+T_2 R_2+R_2 T_2^\intercal
%+T_1 R_3+R_3 T_1^\intercal)\\
%\nonumber
%&+\frac{1}{3}
%(T_1 T_2 R_1+R_1 T_2^\intercal T_1^\intercal
%+T_2 T_1 R_1+R_1 T_1^\intercal T_2^\intercal\\
%\nonumber
%&+T_1^2 R_2+R_2 T_1^\intercal{}^2)
%+\frac{1}{6}
%(T_1 R_1 T_2^\intercal
%+T_2 R_1 T_1^\intercal
%+T_1 R_2 T_1^\intercal)\\
%\nonumber
%&-\frac{1}{4}
%(T_1^3 R_1+R_1 T_1^\intercal{}^3)
%-\frac{1}{12}
%(T_1^2 R_1 T_1^\intercal
%+T_1 R_1 T_1^\intercal{}^2).    
\end{align}
Higher order terms in these series can be calculated but are not discussed in this paper.

\section{Gaussian ancillary bombardment}\label{AncillaryBombardmentGQM}
In this section we construct the Gaussian channel corresponding to a specific physically motivated situation that we refer to as \textit{Gaussian ancillary bombardment}, in analogy with the ancillary bombardment introduced in \cite{Grimmer2016a}. Following this we use the results of the previous section to calculate the interpolation generators and expand them as a series in $\delta t$. Finally, we will analyze these expansions order by order using the partition developed in \cite{ArXivGrimmer2017b}.

In a general Gaussian ancillary bombardment scenario, we consider a Gaussian system that is repeatedly bombarded by a series of Gaussian ancillae. Updating the system's state via \eqref{UpdateSchemeXX} and \eqref{UpdateSchemeVV} here corresponds to the system  interacting with one of these Gaussian ancillae. An intuitive example of such a scenario (and one we analyze in Sec \ref{Example}) is a harmonic oscillator bombarded by a thermal bath of harmonic oscillators.

Let us consider a system, $\text{S}$, to be a Gaussian system composed of $N_\text{S}$ modes. Likewise let each ancilla, $\text{A}$, be a Gaussian system composed of $N_\text{A}$ modes. Together they form a joint system, $\text{SA}$, which is Gaussian and is composed of $N_\text{S}+N_\text{A}$ modes. Note that dimensions of $\text{S}$, $\text{A}$ and $\text{SA}$'s phase spaces are $2N_\text{S}$, $2N_\text{A}$, and $2N_\text{S}+2N_\text{A}$ respectively.

The system and ancilla's quadrature operators are collected together into the operator vector
\be
\hat{\bm{X}}_\text{SA}=(\hat{\bm{X}}_\text{S},\hat{\bm{X}}_\text{A})^\intercal.
\ee
Since the system's and ancilla's observables live in different Hilbert spaces, all pairs of their observables commute with each other. Thus they have the joint symplectic form,
\be
\Omega_\text{SA}=
\begin{pmatrix}
\Omega_\text{S} & 0\\
0 & \Omega_\text{A}
\end{pmatrix}
\ee
where $\Omega_\text{S}$ and $\Omega_\text{A}$ are the symplectic forms in the phase space of S and A respectively. 

We assume that the system and ancilla are initially uncorrelated, having the initial joint mean vector,
\be
\bm{X}_\text{SA}(0)=(\bm{X}_\text{S}(0),\bm{X}_\text{A}(0))^\intercal,
\ee
and the initial joint covariance matrix,
\be
\sigma_\text{SA}(0)=
\begin{pmatrix}
\sigma_\text{S}(0) & 0\\
0 & \sigma_\text{A}(0)
\end{pmatrix}.
\ee
Further we assume that they evolve under a quadratic Hamiltonian,
\bel{HSADef}
\hat{H}_\text{SA}
=\frac{1}{2}\hat{\bm{X}}_\text{SA}^\intercal  \, F_\text{SA} \, \hat{\bm{X}}_\text{SA}
+\bm{\alpha}^\intercal_\text{SA}\hat{\bm{X}}_\text{SA}
\ee
where $F_\text{SA}$ is real and symmetric and $\bm{\alpha}_\text{SA}$ is real. 

It is useful to divide this Hamiltonian into subblocks corresponding to the system and ancilla's phase spaces as,
\be
F_\text{SA}=
\begin{pmatrix}
F_\text{S} & G\\
G^\intercal & F_\text{A}
\end{pmatrix},
\quad \quad
\bm{\alpha}_\text{SA}=
\begin{pmatrix}
\bm{\alpha}_\text{S}\\
\bm{\alpha}_\text{A}
\end{pmatrix}.
\ee
Note that $F_\text{S}$ and $F_\text{A}$ are symmetric and that $G$ is not generally square, having dimensions $2 N_\text{S}$ by $2 N_\text{A}$. 

Divided this way we can see that $F_S$ and $\bm{\alpha}_\text{S}$ correspond to the system's free Hamiltonian,
\be
\hat{H}_\text{S}
=\frac{1}{2}\hat{\bm{X}}_\text{S}^\intercal  \, F_\text{S} \, \hat{\bm{X}}_\text{S}
+\bm{\alpha}^\intercal_\text{S}\hat{\bm{X}}_\text{S}.
\ee
Similarly $F_\text{A}$ and $\bm{\alpha}_\text{A}$ correspond to the ancilla's free Hamiltonian,
\be
\hat{H}_\text{A}
=\frac{1}{2}\hat{\bm{X}}_A^\intercal  \, F_\text{A} \, \hat{\bm{X}}_\text{A}
+\bm{\alpha}^\intercal_\text{A}\hat{\bm{X}}_\text{A}.
\ee
Finally, we can see that the $G$ matrix contains all of the couplings between the system and the ancilla, corresponding to the interaction Hamiltonian, 
\be
\hat{H}_\text{I}
=\frac{1}{2}\hat{\bm{X}}_\text{S}^\intercal  \, G \, \hat{\bm{X}}_\text{A}
+\frac{1}{2}\hat{\bm{X}}_\text{A}^\intercal  \, G^ \intercal\, \hat{\bm{X}}_\text{S}.
\ee

Next we compute the effect that evolving for a time $\delta t$ under this Hamiltonian has on the system (determining $T(\delta t)$, $\bm{d}(\delta t)$, and $R(\delta t)$). In order to do this we compute the evolution of the joint system then isolate the effect on the system. This evolution is unitary and therefore given by a symplectic-affine transformation in the joint phase space. Specifically,
\begin{align}
\label{SXUpdt}
\bm{X}_\text{SA}(\delta t)
&=S_\text{SA}(\delta t) \, \bm{X}_\text{SA}(0)+\bm{d}_\text{SA}(\delta t),\\
\label{SVUpdt}
\sigma_\text{SA}(\delta t)
&=S_\text{SA}(\delta t) \, \sigma_\text{SA}(0) \, S^\intercal_\text{SA}(\delta t)
\end{align}
where
\begin{align}
\label{AppSHamDef}
S_\text{SA}(\delta t)&=\text{exp}(\Omega_\text{SA} F_\text{SA} \, \delta t),\\
\label{AppdHamDef}
\bm{d}_\text{SA}(\delta t)&=\frac{\text{exp}(\Omega_\text{SA} F_\text{SA} \, \delta t)-\openone_{2 N_\text{S}+2 N_\text{A}}}{\Omega_\text{SA} F_\text{SA}} \, \Omega_\text{SA} \, \bm{\alpha}_\text{SA}.
\end{align}
In order to find the effective update on the system's state we can divide these into blocks as
\be
\nonumber
S_\text{SA}(\delta t)
=\begin{pmatrix}
M_\text{SS}(\delta t) & M_\text{SA}(\delta t) \\
M_\text{AS}(\delta t) & M_\text{AA}(\delta t) \\
\end{pmatrix}
\ \ \text{and} \ \ 
\bm{d}_\text{SA}(\delta t)
=\begin{pmatrix}
\bm{d}_\text{S}(\delta t) \\ \bm{d}_\text{A}(\delta t) \\
\end{pmatrix}.
\ee
Expanding \eqref{SXUpdt} and \eqref{SVUpdt} over the direct sum between the system and ancilla's phase spaces, one can identify that the reduced state of the system ($\bm{X}_\text{S}$ and $\sigma_\text{S}$) is updated as
\begin{align}
\bm{X}_\text{S}(\delta t)
&=T(\delta t) \, \bm{X}_\text{S}(0)
+\bm{d}(\delta t),\\
\sigma_\text{S}(\delta t)
&=T(\delta t) \, \sigma_\text{S}( 0) \, T^\intercal(\delta t)
+R(\delta t),
\end{align}
where 
\begin{align}\label{TSMdSDef}
T(\delta t)&=M_\text{SS}(\delta t),\\
\bm{d}(\delta t)&=M_\text{SA}(\delta t) \ \bm{X}_\text{A}(0)+\bm{d}_\text{S}(\delta t),\\
R(\delta t)&=M_\text{SA}(\delta t) \, \sigma_\text{A}(0) \, M^\intercal_\text{SA}(\delta t).
\end{align}

With some effort, these can be expanded as a series in $\delta t$ (as in \eqref{TSeries}, \eqref{dSeries}, and \eqref{RSeries}). Using the results of the previous section, we can then write the interpolation generators $A_{\delta t}$, $\bm{b}_{\delta t}$, and $C_{\delta t}$ as a series in $\delta t$ (as in \eqref{ASeries}, \eqref{bSeries}, and \eqref{CSeries}) now with coefficients written explicitly in terms of the Hamiltonian \eqref{HSADef}.

This calculation is tedious but ultimately straightforward. For the first few terms of the expansion of $A_{\delta t}$ it yields
\begin{align}
A_0=&F_S,\\
\label{A1DefHam}
A_1=&\frac{1}{2}G \, \Omega_A G^\intercal,\\
A_2=&-\frac{1}{12} G \, \Omega_A G^\intercal \Omega_S F_S
-\frac{1}{12} F_S \Omega_S G \, \Omega_A  G^\intercal\\
\nonumber
&+\frac{1}{6} G \, \Omega_A F_A \Omega_A G^\intercal.
%\nonumber
%A_3=&-\frac{1}{24} G \, \Omega_A  F_A \Omega_A  G^\intercal \Omega_S F_S
%-\frac{1}{24} F_S\Omega_S G \, \Omega_A  F_A \Omega_A  G^\intercal\\
%\nonumber
%&+\frac{1}{24} F_S\Omega_S G \, \Omega_A  G^\intercal \Omega_S F_S
%+\frac{1}{24} G \, \Omega_A  F_A \Omega_A  F_A \Omega_A  G^\intercal\\
%&-\frac{1}{12} G \, \Omega_A  G^\intercal \Omega_S G \, \Omega_A  G^\intercal.
\end{align}
For the first few terms of the expansion of $\bm{b}_{\delta t}$ we find
\begin{align}
\bm{b}_0&=
\bm{\alpha}_\text{S}
+G\bm{X}_\text{A}(0),\\
\bm{b}_1&=
\frac{1}{2} G \, \Omega_\text{A} F_\text{A}\bm{X}_\text{A}(0)
+\frac{1}{2} G \, \Omega_\text{A}\bm{\alpha}_\text{A},\\ 
\bm{b}_2&=-\frac{1}{12} F_\text{S} \Omega_\text{S} G \, \Omega_\text{A} \bm{\alpha}_\text{A}
+\frac{1}{6} \Omega_\text{S} G \, \Omega_\text{A} F_\text{A} \Omega_\text{A} \bm{\alpha}_\text{A}\\
\nonumber
&-\frac{1}{12} F_\text{S} \Omega_\text{S} G \, \Omega_\text{A} F_\text{A} \bm{X}_\text{A}(0)
+\frac{1}{6} G \, \Omega_\text{A} F_\text{A} \Omega_\text{A} F_\text{A} \bm{X}_\text{A}(0)\\
\nonumber
&-\frac{1}{12} G \, \Omega_\text{A} G^\intercal \Omega_\text{S} \bm{\alpha}_\text{S}
-\frac{1}{12} G \, \Omega_\text{A} G^\intercal \Omega_\text{S} G \bm{X}_\text{A}(0).
%\bm{b}_3&=\frac{1}{24} G \, \Omega_A F_A \Omega_A F_A \Omega_A \bm{\alpha}_A
%-\frac{1}{24} F_S\Omega_S G \, \Omega_A F_A \Omega_A \bm{\alpha}_A\\
%\nonumber
%&-\frac{1}{24} G \, \Omega_A F_A \Omega_A G^\intercal \Omega_S \bm{\alpha}_S
%+\frac{1}{24} F_S\Omega_S G \, \Omega_A G^\intercal \Omega_S \bm{\alpha}_S\\
%\nonumber
%&-\frac{1}{24} G \, \Omega_A F_A \Omega_A G^\intercal \Omega_S G \bm{X}_A(0)
%-\frac{1}{12} G \, \Omega_A G^\intercal \Omega_S G \, \Omega_A F_A \bm{X}_A(0)\\
%\nonumber
%&+\frac{1}{24} F_S\Omega_S G \, \Omega_A G^\intercal \Omega_S G \bm{X}_A(0)
%-\frac{1}{24} F_S\Omega_S G \, \Omega_A F_A \Omega_A F_A \bm{X}_A(0)\\
%\nonumber
%&+\frac{1}{24} G \, \Omega_A F_A \Omega_A F_A \Omega_A F_A \bm{X}_A(0)
%-\frac{1}{12} G \, \Omega_A G^\intercal \Omega_S G \, \Omega_A \bm{\alpha}_A.
\end{align}
Finally, the first few terms of the expansion of $C_{\delta t}$ are
\begin{align}
C_0&=0,\\
\label{C1DefHam}
C_1&=\Omega_\text{S} G \sigma_\text{A}(0) G^\intercal \Omega^\intercal_\text{S},\\
\label{C2DefHam}
C_2&=\frac{1}{2} \Omega_\text{S} G \big(\Omega_\text{A} F_\text{A} \sigma_\text{A}(0)+\sigma_\text{A}(0) (\Omega_\text{A} F_\text{A})^\intercal\big) G^\intercal \Omega^\intercal_\text{S}.
%C_3&=\frac{1}{6} \Omega_S G\sigma_A(0) F_A \Omega_A{}^\intercal F_A \Omega_A{}^\intercal G^\intercal \Omega_S{}^\intercal\\
%\nonumber
%&+\frac{1}{3} \Omega_S G \, \Omega_A F_A \sigma_A(0) F_A \Omega_A{}^\intercal G^\intercal \Omega_S{}^\intercal\\
%\nonumber
%&+\frac{1}{6} \Omega_S G \, \Omega_A F_A \Omega_A F_A \sigma_A(0) G^\intercal \Omega_S{}^\intercal\\
%\nonumber
%&-\frac{1}{12} \Omega_S G \Omega_A G^\intercal \Omega_S G\sigma_A(0) G^\intercal \Omega_S{}^\intercal\\
%\nonumber
%&-\frac{1}{12} \Omega_S G\sigma_A(0) G^\intercal \Omega_S{}^\intercal G \Omega_A{}^\intercal G^\intercal \Omega_S{}^\intercal\\
%\nonumber
%&-\frac{1}{12} \Omega_S G\sigma_A(0) F_A \Omega_A{}^\intercal G^\intercal \Omega_S{}^\intercal F_S \Omega_S{}^\intercal\\
%\nonumber
%&-\frac{1}{12} \Omega_S F_S \Omega_S G \Omega_A F_A \sigma_A(0) G^\intercal \Omega_S{}^\intercal\\
%\nonumber
%&-\frac{1}{12} \Omega_S G \, \Omega_A F_A \sigma_A(0) F_A \Omega_A{}^\intercal G^\intercal \Omega_S{}^\intercal\\
%\nonumber
%&-\frac{1}{12} \Omega_S F_S \Omega_S G\sigma_A(0) G^\intercal \Omega_S{}^\intercal F_S \Omega_S{}^\intercal.
\end{align}

It is worth noting the functional dependence of $A_{\delta t}$, $\bm{b}_{\delta t}$, and $C_{\delta t}$ on the parameters of the bombardment scenario. These include the system free Hamiltonian ($F_\text{S}$ and $\bm{\alpha}_\text{S}$), ancillae free Hamiltonian ($F_\text{A}$ and $\bm{\alpha}_\text{A}$), the interaction Hamiltonian ($G$) and the initial state of the ancilla ($\bm{X}_\text{A}$ and $\sigma_\text{A}$). The interpolation generators depend on these (even non-perturbatively) as
\begin{align}
&A_{\delta t}(F_\text{S},F_\text{A},G),\\
&\bm{b}_{\delta t}(F_\text{S},F_\text{A},G,\bm{\alpha}_\text{S},\bm{\alpha}_\text{A},\bm{X}_\text{A}(0)),\\
&C_{\delta t}(F_\text{S},F_\text{A},G,\sigma_\text{A}(0)).
\end{align}

The $A_{\delta t}$ term (which  implements rotation, squeezing, amplifications and relaxation \cite{ArXivGrimmer2017b}) does not depend on either the   linear part of the Hamiltonians nor on the initial ancilla state. This means that the presence and strength of all of these effects is controlled solely by the nature of the coupling to the environment and not by the particular state of the environment. Recall this is true even in the regime of long-time interactions.  

Additionally, since the dynamics of the mean vector is determined entirely by $A_{\delta t}$ and $\bm{b}_{\delta t}$ it is therefore independent of the initial covariance of the ancilla, $\sigma_\text{A}(0)$. 

It is also interesting to note which types of dynamics become available at each order in the series. To do this we use the results of \cite{ArXivGrimmer2017b} which partitions the generators of Gaussian dynamics into 11 parts based on: (a) whether or not the dynamics allows for energy flow between the system and the environment, (b) whether it allows for entanglement to be created between the system and the environment, (c) whether the effect of the dynamics is state-dependent or state-independent and finally (d) whether it mixes different modes together.

The result of applying this partition to the dynamics generated by Gaussian ancillary bombardment is summarized in Table \ref{Table22} (for details see Appendix \ref{AppGBPart}). 

Summarizing this analysis, at zeroth order we have access to all the types of dynamics present in the system's free Hamiltonian with the option to induce an additional displacement (coming from $\bm{b}_0$). At higher orders the dynamics will generically be able to access all types of displacement and noise. Past zeroth order, the rotation, squeezing and amplification effects (coming from $A$)  that are available to the system alternate between unitary and non-unitary.

\begin{table*}
\begin{tabular}{||c|c|c|c|c|c|c||}
\hline Type of Dynamics & \quad 0th (Free) \quad &  \quad 0th (Induced) \quad & \quad Odd ( $\geq$ 1st) \quad & \quad Even ( $\geq$ 2nd) \quad \\

\hline Single-mode Rotation &        Yes & No &         No &        Yes \\

\hline Single-mode Squeezing &      Yes &    No &         No &        Yes \\

\hline Displacement & Yes &  Yes &        Yes &        Yes \\

\hline Single-mode Squeezed Noise &   No &       No &        Yes &        Yes \\

\hline Amplification/Relaxation &   No &       No &        Yes &         No \\

\hline Thermal Noise &      No &    No &        Yes &        Yes \\

\hline Multi Mode Rotation &   Yes &       No &         No &        Yes \\

\hline Multi Mode Squeezing &   Yes &       No &         No &        Yes \\

\hline Multi Mode Counter-Rotation &   No &       No &        Yes &         No \\

\hline Multi Mode Noise &   No &       No &        Yes &        Yes \\

\hline Multi Mode Counter-Squeezing &   No &       No &        Yes &         No \\
\hline
\end{tabular}  
\caption{The dynamics available to a bombarded Gaussian system at each order in $\delta t$. The eleven types of dynamics listed in this table are described in detail in \cite{ArXivGrimmer2017b}. The zeroth order effects are further divided into those available through the system's free Hamiltonian and those which can be induced through the interaction.
}\label{Table22}
\end{table*}

Finally, before analyzing each of these expansions order by order, we make some comments about when open Gaussian dynamics in general, and Gaussian ancillary bombardment in particular, can lead to purification. This is an important characterization because dynamics being able to increase the purity of at least one state is a prerequisite for the dynamics to be able to capture the process of thermalization (e.g. cooling through bombardment by a cold environment).

Following \cite{Grimmer2017a} we say that a map can purify if there exists a state whose purity increases under the map. The purity of a Gaussian state \cite{GPurity} is given in our notation by
\be
\mathcal{P}=\text{Tr}(\rho^2)
=\frac{1}{\text{det}(\sigma)}.
\ee
A necessary and sufficient condition for Gaussian dynamics to be able to purify is
\bel{GaussianNandS}
\text{Tr}\big(\Omega A\big)<0.
\ee

Within the partition described in \cite{ArXivGrimmer2017b}, only the Gaussian dynamics including amplification/purification effects are capable or purifying. From Table \ref{Table22} we can see that such effects are only available at odd orders. Thus if no purification effects are present at first order, the leading order purification effects will be at third order, generically two orders lower than the leading order noise term, $C_1$, with which they will compete. In subsection \ref{FirstOrderGaussian} we find that many commonly used interaction Hamiltonians cannot purify at first order.

\subsection{Zeroth Order Dynamics}
The zeroth order dynamics (i.e, in the continuum limit, as $\delta t\to 0$) is unitary, since $A_0$ is symmetric and $C_0$ vanishes. Specifically in zeroth order we have the dynamics,
\begin{align}\label{GQMInterpMasterEqs}
\bm{X}_{S}'(t)
&=\Omega(F_{S} \, \bm{X}_S(t)
+\alpha_S +G \, \bm{X}_S(0))\\
\sigma_S'(t)
&=(\Omega F_{S}) \, \sigma_S(t)
+\sigma_S(t) \, (\Omega F_{S})^\intercal.
\end{align}
Comparing this to \eqref{SymplecticDiffXUpHam} and \eqref{SymplecticDiffVUpHam} we can see that this is just evolution under the effective Hamiltonian
\begin{align}\label{Heff0Gaussian}
\hat{H}_\text{eff}^{(0)}
&=\frac{1}{2}\hat{\bm{X}}_S^\intercal  \, F_{S} \, \hat{\bm{X}}_S
+\hat{\bm{X}}_S^\intercal(\bm{\alpha}_S
+ G\bm{X}_A(0))\\
\nonumber
&=\hat{H}_S+\hat{\bm{X}}_S{}^\intercal G\bm{X}_A(0).
\end{align}
This is in line with the general result from \cite{Layden:2015b} showing that rapid repeated interaction (even in a non-Gaussian setting) produces unitary dynamics in the continuum limit. 
In \cite{Layden:2015b} this result was interpreted as saying that in this regime the ancillae affect the system but do not entangle with it (they ``push'' the system, but do not``talk'' to it). Further it was shown in \cite{Layden:2015b} that by switching evolution between (non-commuting) $\hat{H}_\text{S}$ and $\hat{H}_\text{eff}^{(0)}$ one can generally gain full unitary control of the system. However this cannot be done within the context of Gaussian ancillary bombardment. 

In fact we will argue that only a limited range of Gaussian dynamics is available to the system at zeroth order. Specifically, unlike in \cite{Layden:2015b}, by turning on and off the environment, one can only adjust the system's Hamiltonian by a linear term in $\hat{\bm{X}}_\text{S}$, as can be seen from \eqref{Heff0Gaussian}. Such a modification of the system's Hamiltonian can only apply a displacement and cannot affect the dynamics of the system's covariance matrix. Thus while we are able to push the Gaussian state around as we like in phase space, we are not able to adjust its ``shape'' at will. 

Finally, for completeness we note that since the zeroth order evolution is unitary it is trivially completely positive. Explicity from \eqref{DiffCPCond},
\bel{CPCheck0}
C_0=0\geq\ii\Omega_S (A_0-A_0^\intercal)\Omega_S^\intercal=0.
\ee

\subsection{First Order Dynamics}\label{FirstOrderGaussian}
At first order, we see a new displacement term (from $b_1$), the first noise in the dynamics (from $C_1$) and several other non-unitary effects (from $A_1$). Specifically, from Table \ref{Table22} we can see that in addition to the displacement effects coming from $b_1$ we can have all three kinds of noise (from $C_1$) as well as amplification/relaxation, multi-mode counter-rotation, and counter-squeezing coming from $A_1$. Note that we do not have access to single or multi-mode rotation or squeezing at first order. Since noise is generically present at first order (see below) single or multi-mode rotation or squeezing will be generally be subleading to the noise in the dynamics. 

At this order the dynamics coming from both $A_1$ and $C_1$ is non-unitary ($A_1$ is antisymmetric, and noise is always non-unitary), thus the only unitary effects at first order come from $\bm{b}_1$. These effects give a first order correction to the effective Hamiltonian 
\be
\hat{H}_\text{eff}
=\hat{H}_\text{eff}^{(0)}
+\delta t \, \hat{H}_\text{eff}^{(1)}
+\mathcal{O}(\delta t^2)
\ee
of
\be
\hat{H}_\text{eff}^{(1)}
=\hat{\bm{X}}_\text{S}^\intercal \, \bm{b}_1
=\frac{1}{2}\hat{\bm{X}}^\intercal_\text{S} 
G \, \Omega_A 
\big(F_A\bm{X}_A(0)
+\bm{\alpha}_A\big).
\ee
This correction can be understood as accounting for the ancilla freely evolving during the interaction.

The first order noise term is given by
\be
C_1=\Omega_S G \, \sigma_A(0) \, G^\intercal \Omega_S^\intercal
\ee
which we note is positive semi-definite \mbox{($C_1\geq 0$)}, since \mbox{$\sigma_\text{A}(0)\geq0$}. This noise vanishes only if $G=0$ (there is no interaction) or if $\sigma_A(0)$ is singular (i.e., infinitely squeezed) and  $G^\intercal\Omega_S^\intercal$ maps entirely into the kernel of $\sigma_A(0)$. 

As discussed above, a necessary and sufficient condition for Gaussian dynamics to cause purification is \eqref{GaussianNandS}. Since the zeroth order dynamics is unitary the first opportunity for purification is at first order. This can happen if and only if
\bel{FirstOrderPurifyNandS}
0>\text{Tr}\big(\Omega_\text{S} A_1\big)
= \frac{1}{2} \text{Tr}\big(\Omega_\text{S} \, G \,  \Omega_\text{A} \,  G^\intercal\big).
\ee

In \cite{Grimmer2016a} a necessary and sufficient condition for dynamics causing causing purification at leading order was given in a general (non-Gaussian) ancillary bombardment scenario provided the system is finite dimensional. As such the results described there cannot be applied to Gaussian systems. They concluded that in order to cause purification at leading possible order an interaction must be ``sufficiently complicated''. In particular they found that a tensor product interaction Hamiltonian of the form
\be
H_\text{I}=\hat{Q}_\text{S}\otimes \hat{R}_\text{A}
\ee
will not purify at leading order. We will now prove that this result in fact does extend to the Gaussian context despite the infinite dimensional nature of the systems and ancillae.

Both $\hat{Q}_\text{S}$ and $\hat{R}_\text{A}$ must be linear in their respective quadrature operators, and so
\be
\hat{Q}_\text{S}
=\bm{u}^\intercal\hat{\bm{X}}_\text{S}
=\hat{\bm{X}}_\text{S}^\intercal\bm{u}
\quad\text{and}\quad
\hat{R}_\text{A}
=\bm{v}^\intercal\hat{\bm{X}}_\text{A}
=\hat{\bm{X}}_\text{A}^\intercal\bm{v}
\ee
for some real vectors $\bm{u}$ and $\bm{v}$ in order that $H_\text{I}$  be quadratic in these operators. Thus we can write
\be
\hat{H}_\text{I}
=\frac{1}{2}\hat{\bm{X}}_\text{S}^\intercal  \, G \, \hat{\bm{X}}_\text{A}
+\frac{1}{2}\hat{\bm{X}}_\text{A}^\intercal  \, G^ \intercal\, \hat{\bm{X}}_\text{S}.
\ee
with
\be
G=\bm{u}\bm{v}^\intercal.
\ee
Thus in Gaussian quantum mechanics, tensor product interactions correspond to rank one interaction matrices.

From \eqref{FirstOrderPurifyNandS} we can quickly see that a rank one interaction cannot purify at leading order since
\begin{align}
\text{Tr}\big(\Omega_\text{S} G \Omega_\text{A} G^\intercal\big)
&=\text{Tr}\big(\Omega_\text{S} \bm{u}\bm{v}^\intercal \Omega_\text{A} \bm{v}\bm{u}^\intercal\big)\\
\nonumber
&= \bm{u}^\intercal\Omega_\text{S} \bm{u} \ \bm{v}^\intercal \Omega_\text{A} \bm{v}\\
\nonumber
&=0
\end{align}
since $\Omega_\text{S}$ and $\Omega_\text{A}$ are antisymmetric.

Thus we have extended the result of \cite{Grimmer2016a} that ``simple'' interaction Hamiltonians cannot cause purification at leading order in rapid bombardment from finite dimensional systems to include Gaussian systems.

Moreover, for rank one interactions, purification will not arise at second order since all effects coming from $A_2$ are unitary. Thus the first purification effects can only arise at third order, generically two orders below the leading order noise terms that  any purification effects would compete with.

\begin{comment}
Hey Dan, weren't we meeting at 3pm today?

Your right, I got caught up adding a new result to the paper. I proved the purification conditions in the Gaussian context. The old result doesnt apply to infinite dimesntional systems but all the results carry over analogously.

ah nice... 

So are you coming before 4pm? that's the time of the next meeting

I don't think I will be on campus today. I have added a bit of content to the paper so we should read over that and send it out tomorrow, i think

okay are you gonna be here on Wednesday?

No I am leaving wednesday

okay, then let me konw when the new content is added and i'll go throughit on my own before sending it and let you know.

If you want I can highlight the new content

Yes please

Ill have it done by tonight

ok, jsut shoot an email when you're done

I kind of like talking this way lol. will do

 lol yeah... well at least one makes sure that the person editing will read it :P
 
 Okay nice job about the proof. Looking forward to reading it.
\end{comment} 

Finally we show that up to first order the dynamics is completely positive. Assuming that the ancillae start in a valid state we have
\be
\sigma_A \geq\ii \, \Omega_A
\ee
Multiplying this by $\Omega_S G$ and $G^\intercal \Omega_S^\intercal$ on either side maintains the inequality, yielding
\be
\Omega_S G \sigma_A G^\intercal \Omega_S^\intercal
\geq \ii \, \Omega_S G \, \Omega_A G^\intercal\Omega_S^\intercal,
\ee
but here we can recognize $C_1$ and $A_1$ from \eqref{A1DefHam} and \eqref{C1DefHam}:
\bel{CPCheck1}
C_1\geq 2\ii \, \Omega_S A_1\Omega_S^\intercal
=\ii \, \Omega_S (A_1-A_1^\intercal)\Omega_S.
\ee
where we have used the antisymmetry of $A_1$. This is exactly the complete positivity condition, \eqref{DiffCPCond}, at first order. Adding this inequality to \eqref{CPCheck0} we confirm the dynamics is completely positive at first order.

\subsection{Second Order Dynamics}

At second order the effective Hamiltonian is 
\be
\hat{H}_\text{eff}
=\hat{H}_\text{eff}^{(0)}
+\delta t \, \hat{H}_\text{eff}^{(1)}
+\delta t^2 \, \hat{H}_\text{eff}^{(2)}
+\mathcal{O}(\delta t^3)
\ee
with
\begin{align}
\hat{H}_\text{eff}^{(2)}
&=\frac{1}{4}\hat{\bm{X}}_S^\intercal  (A_2+A_2^\intercal) \hat{\bm{X}}_S
+\hat{\bm{X}}_S^\intercal \, \bm{b}_2
\end{align}
and there is a further correction coming from both $A_2$ and $\bm{b}_1$.  This is the first order at which we have a correction to the effective Hamiltonian  that is quadratic in the quadrature operators, allowing for single and multi-mode rotations and squeezings.

At second order (and in fact at all even orders) the $A_2$ term does not contribute to the non-unitary dynamics. The only new non-unitary dynamics at this order comes from the new noise term $C_2$. As we can see from \eqref{C2DefHam}, this term can be interpreted as a correction to the $C_1$ noise term accounting for the ancilla's covariance matrix undergoing free evolution during the interaction. 

Up to second order the dynamics is completely positive. Proving this amounts to showing that \eqref{DiffCPCond} is obeyed at second order
\begin{align}\label{CPCheck2}
C_0+\delta t \, C_1+\delta t^2 \, C_2 +\mathcal{O}(\delta t^3) &\geq 
\ii \, \Omega_S (A_0-A^\intercal_0)\Omega_S^\intercal\\
&\nonumber
+\delta t \, \ii \, \Omega_S (A_1-A^\intercal_1)\Omega_S^\intercal\\
&\nonumber
+\delta t^2 \, \ii \, \Omega_S (A_2-A^\intercal_2)\Omega_S^\intercal.
\end{align}
Removing several vanishing terms ($C_0=0, A_0-A^\intercal_0=0$, and $A_2-A^\intercal_2=0$) as well as a factor of $\delta t$ we have
\begin{align}\label{CPCheck2}
C_1+\delta t \, C_2 +\mathcal{O}(\delta t^2) \geq
\ii \, \Omega_S (A_1-A^\intercal_1)\Omega_S^\intercal
\end{align}
In order to prove this we consider the state of the ancilla after it evolves under its free Hamiltonian for a time $\delta t/2$. Since free evolution is a completely positive map, applying it to a valid initial state yields a state that satisfies \eqref{SigmaPosCond}. Computing the covariance matrix of this state to leading order yields,
\begin{align}\label{CPCheck2}
\sigma_A(0)+\frac{\delta t}{2}\big(\Omega_A F_A \sigma_A(0)+\sigma_A(0) (\Omega_A F_A)^\intercal\big)
+\mathcal{O}(\delta t^2)
\geq\ii\Omega_A.
\end{align}
Multiplying by $\Omega_S G$ and $G^\intercal \Omega_S^\intercal$ on the either side and using equation \eqref{A1DefHam}, \eqref{C1DefHam}, and \eqref{C2DefHam}  yields
\begin{align}\label{CPCheck2}
C_1+\delta t C_2 +\mathcal{O}(\delta t^2)\geq 2\ii \, \Omega_S A_1\Omega_S^\intercal
=\ii \, \Omega_S (A_1-A^\intercal_1)\Omega_S^\intercal
\end{align}
where in the last step we again employed the antisymmetry of $A_1$. This is the desired result.

\subsection{Higher Order Dynamics}
At third and higher orders the dynamics of the interpolation scheme is not always completely positive. This could indicate either the presence of non-Markovianity (specifically RHP non-Markovianity   \cite{RHPnonMarkov}) in the interpolated dynamics or the breakdown of one of the assumptions underlying the construction of the interpolation scheme, for instance the time-independence of the interpolation generators.

Note that while the differential dynamics given by \eqref{GQMInterpMasterEqsX} and \eqref{GQMInterpMasterEqsV} may not be completely positive, the discrete dynamics described by \eqref{UpdateSchemeXX} and \eqref{UpdateSchemeVV} is guaranteed to be completely positive at every time step (i.e. when $t=n\, \delta t$) since the interpolated dynamics matches the discrete dynamics at those precise times. In the language of \cite{Layden:2015b,Grimmer2016a} this error is termed stroboscopic and can be bounded by a combination of the timescale $\delta t$ and the energy scale of the dynamics, $E$.

\section{Thermalization of a Harmonic Oscillator}\label{Example}
As a first relevant physical scenario that Gaussian ancillary bombardment can shed some light on, we consider the analysis of the time evolution of a harmonic oscillator subject to short interactions with the components of a thermal reservoir. This is a picture usually associated with thermalization processes and as such we would a-priori expect that this evolution has fixed points related to the second law of thermodynamics.

More concretely, one might expect in such a scenario that the harmonic oscillator will thermalize to the temperature of the reservoir, in a way largely independent of the coupling between them. Perhaps surprisingly, we will show that the system does not always thermalize. Moreover, when it does thermalize its final temperature depends critically on the nature of the coupling to the bath (as well as the bath's temperature as expected).

Let us consider a single harmonic oscillator (the system, S) repeatedly interacting with a series of other harmonic oscillators (the ancillae, A) in thermal states with a fixed temperature.

At this point it is convenient to introduce the following basis for 2 by 2 matrices:
\bel{2by2basis}
\openone_2
=\begin{pmatrix}
1 & 0 \\
0 & 1 \\
\end{pmatrix},
\,
\omega=\begin{pmatrix}
0 & 1 \\
-1 & 0 \\
\end{pmatrix},
\, 
X=\begin{pmatrix}
0 & 1 \\
1 & 0 \\
\end{pmatrix},
\,
Z=\begin{pmatrix}
1 & 0 \\
0 & -1 \\
\end{pmatrix}.
\ee

The system's free Hamiltonian is assumed to be
\be
\hat{H}_\text{S}
=\frac{E_\text{S}}{2} (\hat{q}_\text{S}{}^2+\hat{p}_\text{S}{}^2)
=\frac{E_\text{S}}{2}\begin{pmatrix}
\hat{q}_\text{S} & \hat{p}_\text{S}
\end{pmatrix}
\begin{pmatrix}
1 & 0\\
0 & 1
\end{pmatrix}
\begin{pmatrix}
\hat{q}_\text{S}\\
\hat{p}_\text{S}
\end{pmatrix},
\ee
where $E_\text{S}$ is the energy gap of the oscillator. This Hamiltonian is represented in phase space as
\be
F_\text{S}=E_\text{S}
\begin{pmatrix}
1 & 0\\
0 & 1
\end{pmatrix}
=E_\text{S} \, \openone_2,
\quad\text{and}\quad
\bm{\alpha}_\text{S}=0.
\ee
Similarly the ancillae' free Hamiltonian is assumed to be
\be
\hat{H}_\text{A}
=\frac{E_\text{A}}{2} (\hat{q}_\text{A}{}^2+\hat{p}_\text{A}{}^2)
=\frac{E_\text{A}}{2}\begin{pmatrix}
\hat{q}_\text{A} & \hat{p}_\text{A}
\end{pmatrix}
\begin{pmatrix}
1 & 0\\
0 & 1
\end{pmatrix}
\begin{pmatrix}
\hat{q}_\text{A}\\
\hat{p}_\text{A}
\end{pmatrix},
\ee
where $E_\text{A}$ is the energy gap of the ancilla.
This Hamiltonian is representation in phase space as
\be
F_\text{A}=E_\text{A}
\begin{pmatrix}
1 & 0\\
0 & 1
\end{pmatrix}
=E_\text{A} \, \openone_2
\quad\quad
\bm{\alpha}_\text{A}=0.
\ee
The interaction Hamiltonian between the system and the ancillae is assumed to be a generic quadratic coupling,
\be
\hat{H}_\text{int}
=\frac{1}{2}\hat{\bm{X}}_S^\intercal  \, G \, \hat{\bm{X}}_A
+\frac{1}{2}\hat{\bm{X}}_A^\intercal  \, G^ \intercal\, \hat{\bm{X}}_S
\ee
for any real-valued $2$ by $2$ matrix, $G$. Further, the ancillae are taken to each initially be in the thermal state (see  \cite{GQMRev}),
\be
\sigma_A(0)=\nu_A
\begin{pmatrix}
1 & 0\\
0 & 1
\end{pmatrix}
=\nu_A \, \openone_2,
\quad\quad
\bm{X}_A(0)=0.
\ee
The parameter $\nu$ is a temperature monotone related to the inverse temperature $\beta$ and the energy gap as $E$ as,
\bel{NuBetaRelation}
\nu=\frac{\text{exp}(\beta E)+1}{\text{exp}(\beta E)-1}
\ee
This represents a valid state as long as $\nu_A\geq1$.

As discussed above (and in \cite{Layden:2015b}), at zeroth order the system's dynamics is unitary. In fact, in the Gaussian regime, the dynamics is just the system's free dynamics plus a potential displacement coming from the bombardment. In this case, because the ancilla state has $\bm{X}_A(0)=0$, no new displacement dynamics is induced at zeroth order. Therefore the system evolves freely at zeroth order. All new dynamical effects besides free evolution are higher order, thus  associated with a finite interaction duration.

Explicitly computing the zeroth order interpolation generators one finds
\begin{align}
A_0&=E_{S} \ \openone_2,\\
\bm{b}_0&=0,\\
C_0&=0,
\end{align}
which simply describe the free rotation of the system.

We do however see novel dynamical effects at first order. We find 
\begin{align}
A_1&=\frac{1}{2}G \, \Omega_A G^\intercal=\frac{1}{2}\text{det}(G) \, \omega,\\
\bm{b}_1&=0,\\
C_1&=\nu_A \, \Omega_S \, G \, G^\intercal \, \Omega_S^\intercal
\end{align}
for the first order interpolation generators.
These produce non-unitary dynamics in the system. In particular, using the partition developed in \cite{ArXivGrimmer2017b}, we can see that $A_1$ produces amplification or relaxation depending on the sign of $\text{det}(G)$ at a rate $\sim\delta t \, \text{det}(G)$. Specifically if $\text{det}(G)>0$ the effect of this term (alone) is to exponentially shrink the state's mean vector and covariance matrix towards zero. Alternatively if $\text{det}(G)<0$ this term alone would   push the state's mean vector and covariance matrix to grow exponentially. If $\text{det}(G)=0$ this term has no effect. 

This amplification/relaxation competes with the noise introduced at first order by $C_1$. Generically this will include both thermal noise and squeezed noise. If \mbox{$\text{det}(G)\leq 0$} then both the $A_1$ and $C_1$ terms serve to increase the uncertainty of the state. In this case no fixed point is reached, hence the system does not thermalize. However, if $\text{det}(G)>0$ then the two effects come to an equilibrium that is approximately thermal, as we will show below. 

Explicitly the first order master equation for the covariance matrix is
\begin{align}
\frac{\d}{\d t}\sigma_\text{S}(t)
&=\Omega_\text{S}(A_0+\delta t A_1)\sigma_\text{S}(t)\\
&+\sigma_\text{S}(t)(\Omega_\text{S}(A_0+\delta t A_1))^\intercal
+C_0+\delta t \, C_1.
\end{align}

We can expand the system's covariance matrix over the basis \eqref{2by2basis} as
\begin{align}
\sigma_S(t)
=\nu_S(t)\openone_2
+s_\times(t) X
+s_+(t)Z.
\end{align}
where $\nu_S(t)$ captures the system's temperature and $s_\times(t)$ and $s_+(t)$ capture how the state is squeezed.

In terms of these coefficients the first order master equation for the covariance matrix is
\begin{align}
\frac{\d}{\d t}\nu_S(t)
&=-\delta t \,  \text{det}(G) \, \nu_\text{S}(t)
+ \frac{\delta t}{2} \text{Tr}(G^\intercal G) \, \nu_A\\
\frac{\d}{\d t}s_\times(t)
&=-2 \, E_{\text{S}} \, s_+(t)
-\delta t \, \text{det}(G) \, s_\times(t)\\
&\nonumber
-\frac{\delta t}{2} \text{Tr}(G^\intercal X G) \, \nu_A \\
\frac{\d}{\d t}s_+(t)
&=2 \, E_{\text{S}} \, s_\times(t)
-\delta t \,  \text{det}(G) \, s_+(t)\\
&\nonumber
-\frac{\delta t}{2} \text{Tr}(G^\intercal Z G) \nu_A.
\end{align}

These equations have a fixed point if and only if $\text{det}(G)>0$, in which case the fixed point is attractive. In this case the final state of the system is
\be
\sigma_S(\infty)
=\nu_S(\infty) \, \openone_2
+\mathcal{O}(\delta t)
\ee
with
\be
\nu_S(\infty)=\tilde{\nu}_A\coloneqq\frac{\text{Tr}(G^\intercal G)}{2\, \text{det}(G)}\nu_A
\ee
where $\tilde{\nu}_A$ represents the   effective temperature of the ancilla. The system approaces this state at a rate $\delta t \, \text{det}(G)$. Note that the final   temperature of the system depends on the coupling between the system and environment non-trivially.

At this point, one may wonder if it is possible for the system to become colder than its environment through such a rapid bombardment process. Noting that all $2\times 2$ matrices have 
\bel{Frob2Det}
\text{Tr}(G^\intercal G)\geq 2 \, \text{det}(G),
\ee
we see that the system cannot be cooled to have $\nu_\text{S}(\infty)$ lower than $\nu_\text{A}$,
\bel{NuInequality}
\nu_S(\infty)=\tilde{\nu}_A\geq\nu_A.
\ee
However, this does not mean that system cannot become cooler than its environment. Recall from equation \eqref{NuBetaRelation} that $\nu$ is a monotone function of temperature (in fact, it is a monotone function of $\beta E$). Thus \eqref{NuInequality} implies
\bel{BetaInequality}
\beta_\text{S}(\infty)E_\text{S}\leq E_\text{A}\beta_\text{A}
\ee
or equivalently
\bel{BetaInequality2}
T_\text{S}(\infty)\geq \frac{E_\text{S}}{E_\text{A}}
T_\text{A}.
\ee
If the ancilla has has a larger energy gap than the system the system will be cooled to a temperature below that of the ancillae. 

This appears to be connected to the property of Gaussian passivity, introduced in \cite{Brown}. A quantum state is called Gaussian passive iff there exists no Gaussian unitary that can lower the state's energy. 
In fact, if we assume that \(E_\text{S} < E_\text{A}\), then (\ref{BetaInequality}) is the necessary and sufficient condition for Gaussian passivity. Thus, under the condition \(E_\text{S} < E_\text{A}\), the result of bombardment is to evolve the system such that the joint system-ancilla system is Gaussian passive. However, in the case that the system energy gap is larger than that of the ancilla, this result implies that the joint system becomes explicitely Gaussian non-passive! The energetics of the bombardment steady state therefore depend strongly on the ordering of system and ancilla frequencies. This connection warrents further investigation.

The above inequalities ---\eqref{Frob2Det}, \eqref{NuInequality} and \eqref{BetaInequality}-- are saturated (i.e., we have maximal cooling) only for the following two parameter family of interaction matrices
\bel{PerfectThermalizationGForm}
G=g_{1} \, \openone
+g_{w} \, \omega
=\begin{pmatrix}
g_{1} & g_{w}\\
-g_{w} & g_{1}
\end{pmatrix}
\ee
whose associated  Hamiltonians associated  are
\be
\hat{H}_\text{I}
=g_1(\hat{q}_S\hat{q}_A+\hat{p}_S\hat{p}_A)+g_w(\hat{q}_S\hat{p}_A-\hat{p}_S\hat{q}_A).
\ee

Written in terms of the system and ancillae creation and annihilation operators the maximally cooling interaction Hamiltonians are
\bel{MaxCoolaForm}
\hat{H}_\text{I}
=\begin{pmatrix}
\hat{a}_\text{S} & \hat{a}_\text{S}^\dagger
\end{pmatrix}
\begin{pmatrix}
0 & g\\
g^* & 0
\end{pmatrix}
\begin{pmatrix}
\hat{a}_\text{A} \\ \hat{a}_\text{A}^\dagger
\end{pmatrix}.
\ee
where $g=g_1+\ii g_w$. Notice that these are exactly the interaction Hamiltonians that result from dropping all the $\hat{a}_\text{S} \, \hat{a}_\text{A}$ and $\hat{a}_\text{S}^\dagger \, \hat{a}_\text{A}^\dagger$ terms as one does in the rotating wave approximation. Thus taking the rotating wave approximation can have significant phenomenological effects in rapid repeated interaction scenarios. For instance $H_\text{I}=\lambda \,  \hat{q}_S \, \hat{q}_A$ does not thermalize (since it has $\text{det}(G)=0$) but under the rotating wave approximation it causes maximal cooling.
\\~\\
In order to see why the interaction Hamiltonians given by \eqref{MaxCoolaForm} cause the system to equilibrate with its environment it is useful to look at their effect on definite number states. For instance
\begin{align}
\hat{H}_\text{I}\ket{n_S, \, n_A}
\nonumber
&=\big(
g \, \hat{a}_S\hat{a}_A^\dagger
+g^* \, \hat{a}_S^\dagger\hat{a}_A
\big)\ket{n_S, \, n_A}\\
&=g \, \sqrt{n_S}\sqrt{n_A+1}\ket{n_S-1, \, n_A+1}\\
&+g^* \, \sqrt{n_S+1}\sqrt{n_A}\ket{n_S+1, \, n_A-1}
\end{align}
such that the effect of this Hamiltonian is a superposition of either transfering an excitation from S to A or vice versa. In general, these possibilities do not have the same amplitude. If $n_S>n_A$ then
\be
\vert g \, \sqrt{n_S}\sqrt{n_A+1}\vert >
\vert g^* \, \sqrt{n_S+1}\sqrt{n_A}\vert
\ee
such that the amplitude of an excitation being transferred from S to A is larger. Likewise if $n_A>n_S$ then the amplitude of an excitation to be transferred from A to S is larger. Thus this coupling will tend to transfer excitations from the more excited system to the less excited one. As we saw above this ultimately leads to an equilibrium of excitation profiles, $\nu_S=\nu_A$. Note that this is not a thermal equilibrium.
\\~\\
On the other hand, the part of the Hamiltonian
\be
\hat{H}_\text{I}
=h \, \hat{a}_S^\dagger\hat{a}_A^\dagger
+h^* \, \hat{a}_S\hat{a}_A
\ee
 that is eliminated by the rotating wave approximation does not lead to equilibration. Its effect on the definite number state is
\begin{align}
\hat{H}_\text{I}\ket{n_S, \, n_A}
&=\big(
h \, \hat{a}_S^\dagger\hat{a}_A^\dagger
+h^* \, \hat{a}_S\hat{a}_A
\big)\ket{n_S, \, n_A}\\
\nonumber
&=h \, \sqrt{n_S+1}\sqrt{n_A+1}\ket{n_S+1, \, n_A+1}\\
\nonumber
&+h^* \, \sqrt{n_S}\sqrt{n_A}\ket{n_S-1, \, n_A-1}.
\end{align}
That is produces a superposition of both oscillators becoming more excited and both becoming less excited. Notice however that for every $n_S$ and $n_A$,
\be
\vert h \, \sqrt{n_S+1}\sqrt{n_A+1}\vert >
\vert h^* \, \sqrt{n_S}\sqrt{n_A}\vert
\ee
such that joint excitation has a larger amplitude than joint de-excitation. This causes the system to increasingly become more and more excited.
\\~\\
Given a general quadratic interaction Hamiltonian
\begin{align}
\hat{H}_\text{I}
=g \, \hat{a}_S\hat{a}_A^\dagger
+g^* \, \hat{a}_S^\dagger\hat{a}_A
+h \, \hat{a}_S^\dagger\hat{a}_A^\dagger
+h^* \, \hat{a}_S\hat{a}_A,
\end{align}
if $\vert h\vert>\vert g\vert$ the system does not equilibrate. However if $\vert g\vert>\vert h\vert$ then the system equilibrates to have
\be
\nu_S(\infty)
=\frac{\text{Tr}(G^\intercal G)}{2\, \text{det}(G)}\nu_A
=\frac{\vert g \vert^2+\vert h \vert^2}{\vert g \vert^2-\vert h \vert^2}\nu_A.
\ee
The final state of the system is determined by a competition between these equilibrating and exciting effects.

\section{Conclusion}
We have considered the dynamics induced in a generic Gaussian system when rapidly bombarded (at a frequency $1/\delta t$) by a series of Gaussian ancillae, a scenario we call \textit{Gaussian ancillary bombardment}. This scenario covers (as a particular case) a harmonic oscillator bombarded by a thermal bath of harmonic oscillators.

%To fully characterize Gaussian ancillary bombardment, we have ported the general formalism of rapid repeated interaction developed in \cite{Layden:2015b,Grimmer2016a,ArXivGrimmer2017b} to Gaussian quantum mechanics, constructing an interpolating master equation for the discrete induced Gaussian dynamics. We then applied these techniques to the particular scenario of Gaussian ancillary bombardment and analyze the resulting dynamics as a series in $\delta t$ and then analyze this dynamics in terms of unitarity, ability to cause energy flow, state-dependence and mode mixing. 

We have applied this formalism to the relevant case of thermalization by interaction with an environment by investigating the particular case of an harmonic oscillator bombarded by the constituents of a thermal bath of harmonic oscillators. 

We have explicitly shown that the equilibration of systems continually bombarded by the micro-constituents of a thermal reservoir is much richer than just the naive expectation that `the system will evolve to reach the environment's temperature'. Namely, we analyzed in depth the effect that the coupling of the system to the ancillae composing the thermal bath have on the systems dynamics. In particular we have exactly characterized the couplings which cause the system to reach a thermal fixed point. Perhaps surprisingly we showed that most couplings will not even equilibrate (e.g. \mbox{$H_\text{I}\sim q_\text{S}\otimes q_\text{E}$}). Furthermore, we analyzed the effect that the nature of the system-environment coupling has on whether the system equilibrates or not and how the final temperature of the system depends on this coupling. Remarkably, we find that in the space of possible couplings only an extremely limited set of interactions causes the system to thermalize to the temperature of its environment. We relate such couplings to the rotative wave approximation.

We have found other more general results that apply to Gaussian ancillary bombardment. For example we found that a sufficiently complicated interaction Hamiltonian is required to cause purification in this context. We also found that in a general Gaussian Bombardment scenario the presence and strength of any dynamics implementing rotation, squeezing and amplification are entirely independent of the state of the ancillae  constituting the environment (even outside perturbation theory).

Expanding the dynamics as a series in $\delta t$ we found that different types of dynamics are available at each order in the inverse of the interaction frequency with the following consequences: (a) at zeroth order the evolution is unitary as predicted by the general results in \cite{Layden:2015b};  (b) however, unlike in \cite{Layden:2015b} in the Gaussian regime only a limited range of dynamics (only displacements) can be induced in the system at zeroth order; (c) past zeroth order noise and displacement effects are generically present;  (d) rotations, squeezing and amplification effects alternate between unitary and non-unitary at each order.

Our work paves the way to addressing open questions related to the thermodynamics of systems bombarded by environments, and how the energy and information flows between system and environments depend on the particular microscopic details of the interaction.

%We have recovered the general result of \cite{Grimmer2016a} that the interpolated dynamics are completely positive at zeroth and first order, extending this to second order as well in the Gaussian regime.

\acknowledgments

AK, EMM and RBM acknowledge support through the Discovery program of the Natural Sciences and Engineering Research Council of Canada (NSERC). DG acknowledges support by NSERC through the Vanier Scholarship. EGB also acknowledges support by NSERC through their Postdoctoral Fellowship.

\appendix
\section{Constructing the Interpolation Schemes}\label{AppGQMInterpolate}
In this appendix, we construct the interpolation generators described in Sec. \ref{InterpolateGQM} following the technique developed in \cite{Grimmer2016a} and \cite{Grimmer2017a}. 

Specifically we are given a discrete update channel
\begin{align}
\label{AppUpdateSchemeXX}
\bm{X}((n+1)\delta t)
&=T(\delta t) \, \bm{X}(n \, \delta t)
+\bm{d}(\delta t)\\
\label{AppUpdateSchemeVV}
\sigma((n+1)\delta t)
&=T(\delta t) \, \sigma(n \, \delta t) \, T(\delta t)^\intercal
+R(\delta t)
\end{align}
for some $T(\delta t)$, $\bm{d}(\delta t)$, and $R(\delta t)$ with 
\bel{AppNothingNoTime}
T(0)=\openone_{2N}, \ \ \  \bm{d}(0)=0, \ \ \ \text{and} \ \ \ R(0)=0
\ee
and that
\bel{AppFiniteRate}
T'(0), \ \ \  \bm{d}'(0), \ \ \ \text{and} \ \ \ R'(0) \ \ \ \text{exist}
\ee
and finally that $T(\delta t)$ is non-singular.

From the above update we here construct a Gaussian master equation of the general form
\begin{align}
\label{AppGQMInterpMasterEqsX}
\bm{X}'(t)
&=\Omega(A_{\delta t} \, \bm{X}(t)
+\bm{b}_{\delta t})\\
\label{AppGQMInterpMasterEqsV}
\sigma'(t)
&=(\Omega A_{\delta t}) \, \sigma(t)
+\sigma(t) \, (\Omega A_{\delta t})^\intercal
+C_{\delta t}
\end{align}
for some $A_{\delta t}$, $\bm{b}_{\delta t}$, and $C_{\delta t}$ such that solving the master equation gives dynamics matching the discrete update at every time point, $t=n \, \delta t$. 

We find unique interpolation generators ($A_{\delta t}$, $\bm{b}_{\delta t}$, and $C_{\delta t}$) by assuming that they are time-independent and that they converge as $\delta t\to0$.

We will begin by constructing the generators for $\bm{X}$ and then repeat the procedure with the necessary modification for $\sigma$. 

In order to apply the technique developed in \cite{Grimmer2016a} and \cite{Grimmer2017a} we need a linear update equation, not a linear-affine one as we currently have. We can linearize \eqref{AppUpdateSchemeXX} by defining the $2N+1$ dimensional vector
\be
\bm{Y}(n \, \delta t)=
\begin{pmatrix}
1\\\bm{X}(n \, \delta t)
\end{pmatrix}
\ee
which updates as 
\begin{align}\label{UpdateSchemeY}
\bm{Y}((n+1)\delta t)
&=\begin{pmatrix}
1 & \bm{0}^\intercal\\
\bm{d}(\delta t) & T(\delta t)
\end{pmatrix}
\bm{Y}(n\, \delta t)\\
\nonumber
&=\Phi(\delta t)\bm{Y}(n\, \delta t)
\end{align}
where
\be
\Phi(\delta t)
=\begin{pmatrix}
1 & \bm{0}^\intercal\\
\bm{d}(\delta t) & T(\delta t)
\end{pmatrix}.
\ee
The new update matrix, $\Phi(\delta t)$ adopts the regularity properties around $\delta t=0$ that we assumed for $d(\delta t)$ and $T(\delta t)$. Specifically, we have $\Phi(0)=\openone_{2N+1}$, that $\Phi'(0)$ exists and that $\Phi(\delta t)$ is nonsingular.

Solving the recurrence relation \eqref{UpdateSchemeY} we have
\bel{DiscreteMasterEqsy}
\bm{Y}(m \, \delta t)
=\Phi(\delta t)^m\bm{Y}(0).
\ee
In this form we can now apply the formalism developed in \cite{Grimmer2016a} and \cite{Grimmer2017a} to construct an interpolation scheme for $\bm{Y}(t)$ which will in turn will give an interpolation scheme for $\bm{X}(t)$.

Specifically, we find a unique interpolation scheme by making the following three assumptions for the continuous-time evolution: 
\begin{enumerate}
\item The evolution is time-local and time-independent, such that,
\begin{align}
\label{MasterEqsy}
\bm{Y}'(t)
&=\mathcal{L}_{\delta t} \, \bm{Y}(t)
\end{align}
or equivalently,
\begin{align}
\label{IntegratedMasterEqsy}
\bm{Y}(t)
&\coloneqq\exp(t \, \mathcal{L}_{\delta t}) \, \bm{Y}(0),
\end{align}
for some $\mathcal{L}_{\delta t}$.

\item The interpolated evolution exactly matches the discrete dynamics \eqref{DiscreteMasterEqsy} at the end of every time step. Using \eqref{IntegratedMasterEqsy} this means,
\begin{align}
\exp(m \, \delta t \, \mathcal{L}_{\delta t})
=\Phi(\delta t)^m
\end{align}
or equivalently,
\begin{align}
\label{MatchingConditiony}
\exp(\delta t \, \mathcal{L}_{\delta t})
=\Phi(\delta t)
\end{align}

\item The evolution's generator, $\mathcal{L}_{\delta t}$ is well defined in the continuous interaction limit, that is as $\delta t\to0$.
\end{enumerate}

These three conditions uniquely specify the interpolation scheme that is generated by
\begin{align}
\mathcal{L}_{\delta t}
&\coloneqq\frac{1}{\delta t}\text{Log}(\Phi(\delta t))
\end{align}
where we have taken the logarithm's principal branch cut, that is the one with $\text{Log}(\openone_{2N})=0$. The third condition resolves the ambiguity of the logarithm's branch cut by forcing $\text{Log}(\openone_{2N})=0$, which is necessary to make $\mathcal{L}_{\delta t}$ well defined as $\delta t\to0$. Using L'Hopital's rule we see that
\be
\mathcal{L}_0=\Phi'(0)
\ee
which by assumption exists. Finally we require $\Phi(\delta t)$ to be non-singular in order for the logarithm to be evaluated.

Thus we have the master equation for a unique interpolation scheme for the discrete-time evolution, \eqref{DiscreteMasterEqsy}. 
From this we find a master equation for $\bm{X}(t)$ by dividing $\Phi(\delta t)$ into subblocks. Specifically, 
\begin{align}
\mathcal{L}_{\delta t}
&=\frac{1}{\delta t}\text{Log}(\Phi(\delta t))\\
\nonumber
&=\frac{1}{\delta t}\text{Log}
\begin{pmatrix}
1 & \bm{0}^\intercal\\
\bm{d}(\delta t) & T(\delta t)
\end{pmatrix}\\
\nonumber
&=\frac{1}{\delta t}
\begin{pmatrix}
0 & \bm{0}^\intercal\\
\frac{\text{Log}(T(\delta t))}{T(\delta t)-\openone_{2N}}\bm{d}(\delta t) & \text{Log}(T(\delta t))
\end{pmatrix}.
\end{align}
Evaluating \eqref{MasterEqsy} in terms of $\bm{X}(t)$ gives
\begin{align}
\bm{X}'(t)
&=\Omega \, A_{\delta t} \, \bm{X}(t)
+\Omega \, \bm{b}_{\delta t}
\end{align}
with
\bel{AppADef}
\Omega \, A_{\delta t}
=\frac{1}{\delta t}\text{Log}(T(\delta t))
\ee
and
\be
\Omega \, \bm{b}_{\delta t}
=\frac{1}{\delta t}\frac{\text{Log}(T(\delta t))}{T(\delta t)-\openone_{2N}}\bm{d}(\delta t)
\ee
as claimed in section \ref{InterpolateGQM}.

Next we construct the generators for $\sigma$. The process is very similar, except that casting \eqref{AppUpdateSchemeVV} as a linear equation requires an extra step. This entails using
the $\vec$ operation described in Sec \ref{ReviewGQM} (following equation \eqref{OuterToTensor}). Defining
\be
\bm{v}(n \, \delta t)
\coloneqq\vec(\sigma(n \, \delta t))
\ee
we can vectorize the update scheme for $\sigma$
\begin{align}
\sigma((n+1)\delta t)
&=T(\delta t) \, \sigma(n \, \delta t) \, T(\delta t)^\intercal
+R(\delta t)
\end{align}
into one for $\bm{v}$ by using \eqref{VecIdentity}. We find
\begin{align}
\bm{v}((n+1)\delta t)
&=\big(T(\delta t)\otimes T(\delta t)\big) \, \bm{v}(n \, \delta t)
+\vec(R(\delta t)).
\end{align}
This new equivalent update is now formally identical to \eqref{AppUpdateSchemeXX} and applying the same methods as above we can construct a unique interpolation scheme for $\bm{v}=\vec(\sigma)$. Specifically, we find
\be
\bm{v}'(t)
=G_{\delta t} \, \bm{v}(t)
+\bm{h}_{\delta t}
\ee
where 
\begin{align}
G_{\delta t}
&=\frac{1}{\delta t}\text{Log}(T(\delta t) \otimes T(\delta t))
\end{align}
and
\begin{align}
\bm{h}_{\delta t}
&=\frac{1}{\delta t}\frac{\text{Log}(T(\delta t) \otimes T(\delta t)^\intercal)}{T(\delta t) \otimes T(\delta t)^\intercal-\openone_{4N}}\vec(R(\delta t)).
\end{align}
Using the identity
\begin{align}\label{TensorLogSplit}
\text{Log}(T\otimes T)
&=\text{Log}(T)\otimes\openone
+\openone\otimes\text{Log}(T)
\end{align}
and equation \eqref{AppADef} we can simplify $G_{\delta t}$ as
\begin{align}
G_{\delta t}
&=\Omega A_{\delta t}\otimes\openone_{2N}
+\openone_{2N}\otimes \Omega A_{\delta t}.
\end{align}
Reversing the vectorization we used earlier and using \eqref{VecIdentity} we find that $G_{\delta t}$ acts on $\sigma$ as
\be
(\vec^{-1} G_{\delta t} \, \vec)[\sigma]\\
=\Omega A_{\delta t}\sigma
+\sigma \, (\Omega A_{\delta t})^\intercal.
\ee
Similarly, writing $C_{\delta t}=\vec^{-1} \bm{h}_{\delta t}$ we obtain
\bel{AppCDef}
C_{\delta t}
=\vec^{-1}\Big(\frac{1}{\delta t}\frac{\text{Log}(T(\delta t) \otimes T(\delta t))}{T(\delta t) \otimes T(\delta t)-\openone_{4N}} \, \vec\big(R(\delta t)\big)\Big)
\ee
and so we have 
\begin{align}
\sigma'(t)
&=\Omega A_{\delta t}\sigma
+\sigma \, (\Omega A_{\delta t})^\intercal
+C_{\delta t}
\end{align}
as the master equation for $\sigma$, 
with $A_{\delta t}$ and $C_{\delta t}$ given by \eqref{AppADef} and \eqref{AppCDef} respectively as claimed in Sec. \ref{InterpolateGQM}.

\section{Partitioning}\label{AppGBPart}

In this appendix we determine which types of dynamics (described in \cite{ArXivGrimmer2017b}) become available at each order in the series in $\delta t$.  The generators of open Gaussian dynamics 
can be partitioned 
into 11 parts \cite{ArXivGrimmer2017b} based on the following considerations: (a) if the dynamics allows for energy flow between the system and the environment; (b) if it creates entanglement between the system and the environment; (c) if  the effect of the dynamics is state-dependent or state-independent;  (d) whether it mixes different modes together. 
We establish in this appendix
the results (summarized in Table \ref{Table22}) of applying this partition to the dynamics generated by Gaussian ancillary bombardment.

In \cite{ArXivGrimmer2017b}, in order to separate all these effects, $A$ and $C$ were expanded into $2$ by $2$ blocks (and $\bm{b}$ into $2$ by $1$ blocks). To characterize the dynamics in terms of the above criteria, these blocks were then analyzed in several ways including (for $A$ and $C$) their position on or off the block diagonal and their symmetry properties. 
\bel{App2by2basis}
\openone_2
=\begin{pmatrix}
1 & 0 \\
0 & 1 \\
\end{pmatrix},
\,
\omega=\begin{pmatrix}
0 & 1 \\
-1 & 0 \\
\end{pmatrix},
\, 
X=\begin{pmatrix}
0 & 1 \\
1 & 0 \\
\end{pmatrix},
\,
Z=\begin{pmatrix}
1 & 0 \\
0 & -1 \\
\end{pmatrix}.
\ee

We first note from \cite{ArXivGrimmer2017b} that the $\bm{b}$ term always implements displacement when present. As it can receive a non-zero contribution at every order, we say that displacement is available at every order. Moreover, examining the zeroth order term,  \be
\bm{b}_0=
\bm{\alpha}_\text{S}
+G\bm{X}_\text{A}(0),
\ee
we see that it contains a term coming from the system's free Hamiltonian, $\alpha_\text{S}$, as well as a term induced by the interaction, $G\bm{X}_\text{A}(0)$. Displacement is thus available at zeroth order through the system's free Hamiltonian as well as through effects induced by the interaction. This distinction is noted in Table \ref{Table22}.

Next we recall   \cite{ArXivGrimmer2017b} that the $C$ term implements three types of noise: thermal noise, single-mode squeezed noise and multi-mode noise. Multi-mode noise corresponds to off-block diagonal parts of $C$, and  on-block diagonal components of $C$ can be identified as thermal or squeezed by expanding over the $2\times 2$ basis \eqref{App2by2basis}. Excluding at zeroth order (where the noise vanishes), $C_{\delta t}$ generically has all three types of noise present at every order.

Finally, we analyze $A_{\delta t}$. As discussed in \cite{ArXivGrimmer2017b}, this term implements the various squeezing and rotation effects listed in \ref{Table22} as well as amplification and relaxation. As with $C$, these different types of dynamics are distinguished by expanding $A_{\delta t}$ into $2$ by $2$ blocks and further expanding each block over a certain $2\times 2$ basis, \eqref{2by2basis}. 

With respects to this partition, the contributions to $A_{\delta t}$ at each order are generic except that they alternate being symmetric and antisymmetric at each order.  Within the classification system outlined in \cite{ArXivGrimmer2017b} this corresponds to the dynamics being either unitary or non-unitary (there called symplectic/unsymplectic). Specifically, the parts of $A_{\delta t}$ that are symmetric under transpose give unitary dynamics. These include all the single and multi-mode rotations and squeezings. These dynamics are available at every even order. On the other hand the antisymmetric  parts of $A_{\delta t}$  give non-unitary dynamics. This includes the single and multi-mode counter-rotations and counter-squeezings as well as amplification and relaxation. These dynamics are available at every odd order. (As described in \cite{ArXivGrimmer2017b}, counter-squeezing and counter-rotations are ostensibly like squeezing and rotation except that they do not respect the symplectic form and require a sufficient noise level to be completely positive).

As with $\bm{b}_{\delta t}$, in analyzing $A_{\delta t}$ special attention must be paid to the zeroth order dynamics. Examining the zeroth order term $A_0=F_\text{S}$ we see that it only contains a term coming from the system's free Hamiltonian and has no dependenece on the interaction. Thus while at zeroth order all the unitary dynamics coming from $A_{\delta t}$ are technically available, they are only present if they exist in the system's free Hamiltonian i.e. they cannot be induced. This distinction is noted in Table \ref{Table22}.

\bibliography{references}

\end{document}